\documentclass[12pt]{iopart}

\usepackage{graphicx}  
\begin{document}

\title{Scalar field--perfect fluid correspondence and nonlinear perturbation equations}

\author{Roberto Mainini}

\address{Institute of Theoretical Astrophysics, University of Oslo, Box 1029, 
0315 Oslo, Norway}

\begin{abstract}
The properties of dynamical Dark Energy (DE) and, in particular,
the possibility that it can form or contribute to stable
inhomogeneities, have been widely debated in recent literature,
also in association to a possible coupling between DE and Dark Matter (DM).
In order to clarify this issue, in this paper we present
a general framework for the study of the nonlinear phases of structure
formation, showing the equivalence between two
possible descriptions of DE: a scalar field $\phi$ self--interacting
through a potential $V(\phi)$ and a perfect fluid with an
assigned negative equation of state $w(a)$. This enables us to show that, in the
presence of coupling, the mass of DE quanta may increase
where large DM condensations are present, so that also DE may
partake to the clustering process.
\end{abstract}

\pacs{ }
\maketitle
\section{Introduction}
\label{sec:intro}

An important part of cosmology concerns the large--scale clustering 
of matter in galaxies and galaxy clusters and its statistics. 
The study of Large Scale Structure (LSS) can reveal many aspects 
of the physics of the early Universe as well as 
its matter content during the cosmic history.

Several observations made over recent years, related to a large extension 
to LSS and anisotropies of the Cosmic Microwave Background (CMB) 
as well as the magnitude--redshift relation for type Ia Supernovae \cite{tegmark}, 
have given us a convincing picture of the energy and matter density in 
the Universe.  

Baryonic matter accounts for no more than 30$\%$ of the mass in galaxy 
clusters while the existence of a large clustered component of 
Dark Matter (DM) seems now firmly established, although its nature is 
still unknown. 
However, they 
contribute to the total energy density of the Universe with 
only a few percent and about 25$\%$ respectively.

No more than another few percent could be accounted for by massive neutrinos, 
but only in the most favorable, but unlikely case. According to \cite{oystein}
the total mass of neutrinos cannot exceed the limit of 1.43 $eV$.
A very small part $(10^{-4})$ of the total energy density is due to massless
neutrinos and CMB radiation. 

The model suggested by observations is only viable if the remaining 75$\%$
is ascribed to the so--called Dark Energy (DE) responsible for the present 
day cosmic acceleration. 

Although strongly indicated by the observations, the existence of DE is even more 
puzzling than DM. It can be identified with a cosmological constant 
or with a yet unknown dynamical component with negative pressure. 
On the other hand, its manifestation can be interpreted as a geometrical 
property of the gravity on large scales denoting a failure of 
General Relativity (GR) on those scales (see \cite{copeland} for a review). 
   
Within the context of GR, as alternative to the cosmological constant,
DE is usually described as a self--interacting scalar field or a cosmic fluid
with negative pressure (see \cite{peebles} and references therein). It is usually assumed that density 
perturbations of DE play a negligible role in the structure formation because
of its very small mass $\sim H$ ($H$ being the Hubble parameter). 
Accordingly, perturbations should appear 
only on very large scales and are bound to be linear so that rates of 
structure formation and their growth are influenced by DE only through 
the overall cosmic expansion \cite{mainini}.   

Nevertheless, the clustering properties of dynamical DE have been subject to 
recent debate. The behaviour of DE in the presence of high concentration 
of matter and in particular of DE--matter coupling \cite{interaction,interaction2}, is not clear.

Then, the key question is whether DE actively participates in the clustering 
and virialization processes developing non--linearity on  relevant scales. 
Some attempts to solve the problem have been done in \cite{matarrese,nunes} and an 
analytical expression for uncoupled DE density perturbations valid both in 
the linear and nonlinear regime was derived in \cite{mota}.

The aim of this paper is to provide a general framework which can be useful
for the study of the nonlinear phases of structure formation in cosmologies 
where a coupling between DE and matter is present. The paper is organized as 
follows. 

We assume a Friedmann Robertson Walker (FRW) background Universe. 
Because we are interested in the epoch when structures form, we neglect 
the radiation and consider a system of two coupled 
fluids. A pressureless fluid and one with pressure which can be interpreted 
as DM and DE respectively. Equations for the background evolution are provided
in Sec. \ref{background}.
Then, in Sec. \ref{perturbation}, starting from a general relativistic 
treatment, we derive the Newtonian 
limit of the nonlinear perturbation equations for 
the density contrast.
In sec. \ref{scalarfield}, following \cite{madsen} (see also \cite{bruni}),
 we review
the scalar field--fluid correspondence showing that the formalism described in 
the previous sections apply also to a scalar field. It is then applied to the 
coupled DE model in Sec. \ref{example}. We conclude in Sec. \ref{conclusion}.

\section{Background equations}
\label{background}
Assuming GR holds, let us start from the background equations. 
We adopt the following conventions: the four spacetime coordinates 
are labelled by Greek letters running from 0 to 3 while Latin
letters run from 1 to 3, labelling the spatial coordinates.  
We are interested in the epoch when structures form so that radiation 
is neglected. We consider a Universe filled with baryons ($b$) and 
two coupled perfect fluids with vanishing and non--vanishing  
pressure which, in case, can be interpreted as DM and DE respectively and are 
labelled with $_{dm}$ and $_{de}$. According to general covariance, the sum of the individual stress energy tensors $T^\mu_{(a)\nu}$ ($a=b,dm,de$) must be conserved so we can write:
\begin{equation}
\nabla_{\mu} T^{\mu}_{(b) \nu} = 0\\
\nabla_{\mu} T^{\mu}_{(dm) \nu} = -C_{\nu}\\
\nabla_{\mu} T^{\mu}_{(de) \nu} = C_{\nu}
\label{continuity}
\end{equation}
Here $\nabla_{\mu}$ is the covariant derivative while the 4--vector
$C_\mu$ parametrizes 
a possible interaction between DM and DE being the energy and momentum exchanges  described by 
$C_0$ and $C_i$ respectively. For each component, the 
stress-energy tensor $T^{\mu}_{\nu}$ can be put in the perfect fluid form:
\begin{equation}
T^{\mu}_{\nu} = p \delta^{\mu}_{\nu} + (\rho + p) u^{\mu} u_{\nu}
\label{stressenergy}
\end{equation}
Here  $\rho$ and $p$ are the energy density and  
pressure of the fluid as measured in the rest frame and 
$u^{\mu}$ is its 4--velocity.
We also assume the coupling term 
$C_\nu \equiv C_\nu(\{v_1,...,v_n\}_{dm}, 
\{v_1,...,v_m\}_{de})$ to depend on the 
dynamical variables  of the two coupled fluids that we have generically indicated with 
$v_a (a=1,2,...)$.

In a spatially flat FRW background with metric 
$ds^2 = a^2(\tau) \left( -d\tau^2 + dx^i dx_i \right)$
($\tau$ is the conformal time) the Friedmann 
equation for the scale factor $a$ and equations (\ref{continuity}) read: 

\begin{equation}
 {\cal H}^2 = {8 \pi \over 3} G 
\left(\bar {\rho_b} + \bar {\rho}_{dm} + \bar {\rho}_{de} \right) a^2
\end{equation}
\begin{equation}
\dot {\bar{\rho}}_{b} + 3 {\cal H}\bar {\rho}_{b} = 0\\
\dot {\bar{\rho}}_{dm} + 3 {\cal H}\bar {\rho}_{dm}=  -\bar{C_0}\\
\dot {\bar{\rho}}_{de} + 3 {\cal H}\left(\bar {\rho}_{de} + 
\bar{p}_{de} \right) = \bar{C_0}\\
\label{dens}
\end{equation}
where $~\dot {}~$ denotes the derivative with respect to $\tau$, 
${\cal H}=\dot a/a$ and $~\bar {}~$ denotes the background quantities. 
For a generic function $F(v_1,v_2,...)$ then we have
${\bar F} = F(\bar{v}_1,\bar{v}_2,...)$ so that 
$\bar {C_0} \equiv C_0(\{\bar{v}_1,...\}_{dm},
\{\bar{v}_1,...\}_{de})$. 
Notice that 
from (\ref{continuity}) and (\ref{stressenergy}) it follows that the 
coupling term in the FRW background is $\bar{C_\mu} = (\bar{C_0},0,0,0)$.
\section{Perturbation equations}
\label{perturbation}
According to the standard picture, the observed structures have formed
via gravitational instability from the tiny inhomogeneities  left by cosmic
inflation.

Relativistic theory of cosmological perturbations \cite{ma,kodama} 
provides a useful tool
to understand the evolution of inhomogeneities relating 
the physics of the inflation to CMB anisotropy and LSS.

Despite relativistic effects, the physics of the evolution of perturbations 
is quite simple during the period from the end of inflation to the beginning 
of nonlinear gravitational collapse as they have small amplitudes and 
different wavelengths evolve independently.

Anyway, linear theory breaks down when matter inhomogeneities become 
significantly denser than the cosmic medium. Although the Newtonian theory 
is appropriate to describe the dynamics on scales small compared
to the Hubble radius (but larger than the Schwarzschild radius), full numerical simulations 
are needed to follow the 
nonlinear phases of structure formation (a comprehensive review on numerical techniques is 
provided, e.g, by \cite{bert}). Only in cases where some symmetries 
are present, analytic or semi--analytic schemes apply. 
The simplest approach is to follow a spherical density
enhancement; although real fluctuations are expected to be highly irregular and random,
much work has been done 
along this line, starting with \cite{spheric1,peebles2} who studied the problem
within the frame of pure cold DM models. Then, the results have 
been generalized to the case of $\Lambda$CDM \cite{spheric2} and other DE models 
\cite{mainini,nunes, mainini2}.
In spite of the lack of cogent physical motivations, spherical geometry
yields results suitable to work out useful mass functions and to study their evolutions
within the  Press $\&$ Schechter \cite{PS} -- or similar \cite{sheth} --
approach (see, e.g, \cite{lacey}).

The aim of this section is to provide a general framework which can be used 
to study the nonlinear phases of structure formation in the context of coupled DM--DE
models. Without considering any specific coupling between the dark components we derive and
discuss a set of equations that help to understand the role of the DE and its possible coupling 
with DM in the clustering process.

Dealing with LSS it is sufficient to consider weak gravitational 
fields and nonrelativistic gravitational sources. 
Note that, because of the weak fields, slow motions do not necessarily 
imply small density fluctuations.
 
Starting from a general relativistic treatment, we derive now the perturbation
equations in the Newtonian limit, valid to all orders in the 
density contrast. 
We choose to work in the conformal Newtonian gauge since it 
most closely corresponds to Newtonian gravity and consider only scalar 
perturbations. In this gauge, the metric of a perturbed flat FRW Universe reads:
\begin{equation}
ds^2 = a^2(\tau) \left[ -\left(1 + 2\Phi \right) d\tau^2 + 
\left(1 - 2\Psi \right) dx^i dx_i \right]
\label{metric}
\end{equation}
where $\Phi$ plays the role of the Newtonian potential, $\Psi$ is the 
Newtonian spatial curvature and $|\Phi|,|\Psi| << 1$.

Consider now a generic perfect fluid. In the nonrelativistic limit 
its coordinate velocity $v^i = dx^i/d\tau$ can be treated as a perturbation
of the same order as the metric perturbations. Then, keeping only terms 
linear in $v^i$,$\Phi$ and $\Psi$ and neglecting terms involving products
of  $v^i$ and the metric potentials the components of the 4--velocity 
$u^{\mu} = {dx^\mu / \sqrt {-ds^2}}$ become:
\begin{equation}
u^0 = a^{-1} (1-\Phi)\\ 
u^i = a^{-1} v^i\\ 
u_0 = -a (1+\Phi)\\   
u_i = a v^i \\
\label{4vel}
\end{equation}  
With the same approximations, we can now write the components of the 
stress--energy tensor (\ref{stressenergy}): 
\begin{equation}
T^0_0 = -\rho\\ 
T^0_i = (\rho+p)v^i\\ 
T^i_0 = -(\rho+p)v^i\\
T^i_j = p \delta^i_j\\
\label{tensor}
\end{equation}  
Here $\rho = {\bar\rho}(\tau)+ \delta \rho(\tau,x^i)$ and 
$p = \bar{p}(\tau) + \delta p(\tau,x^i)$
are the perturbed energy density and pressure of the fluid.
Notice that we have not required $\delta \rho$ and $\delta p$ to be small  
compared to the background quantities.

The equations of motion follow from \cite{peebles2}:
\begin{equation}
\nabla_{\mu} T^{\mu}_{\nu} = 
{1\over2}  ~ (\partial_\nu g_{\mu \sigma}) T^{\mu \sigma} - 
{1\over\sqrt{-g}}~\partial_\mu \left(\sqrt{-g} ~T^\mu_\nu \right)= 
  \cal {C}_\nu
\label{conservation}
\end{equation}
where $g$ is the determinant of the metric $g_{\mu \nu}$ given by (\ref{metric})
and to the first order in the 
gravitational potentials $\sqrt{-g} = a^4(1-3\Psi+\Phi)$. 
Possible interactions with 
other components are described by 
${\cal {C}_\mu} = {\bar {\cal C}}_\mu (\tau)+ \delta {\cal {C}}_\mu(\tau,x^i)$
where 
$\delta {\cal C}$ represents deviations from the value $\bar {\cal C}$
taken in an homogeneous and isotropic Universe .
Taking ${\cal C}_\mu = 0, C_\mu, -C_\mu$ equations (\ref{continuity}) are 
then recovered.
From the component $\nu=0$ of (\ref{conservation}) it follows: 
\begin{equation}
{\partial \rho\over \partial \tau}  +
3\left({\cal H}-{\partial \Psi \over \partial \tau} \right)(\rho+p)+
\nabla \cdot \left[(\rho+p){\bf v}\right] = {\cal C}_0
\label{eq1}
\end{equation} 
while the component $\nu=i$ gives:
\begin{equation}
{\partial \over \partial \tau} \left[(\rho+p){\bf v} \right]+
4{\cal H}(\rho+p){\bf v}+\nabla p 
+ (\rho+p)\nabla \Phi = -{\bf {\cal C}}
\label{eq2}
\end{equation} 
($|{\bf v}|^2 = v_i v^i$, $|{\bf {\cal C}}|^2= {\cal C}_i{\cal C}^i$).
Then, making use of the unperturbed part 
of (\ref{conservation}):
\begin{equation} 
\dot {\bar{\rho}} + 3 {\dot a \over a}\left(\bar {\rho} + 
\bar{p} \right) = \bar{\cal{C}}_0
\end{equation}
together with the relation $\rho+p=\bar{\rho}(1+\omega) + \delta\rho
(1+\omega_p)$, where $\omega={\bar p} / {\bar \rho}$
is the equation of state parameter and $\omega_p = \delta p / \delta\rho$,
we can extract from (\ref{eq1}) and (\ref{eq2}) the equations for the 
density contrast $\delta=\delta\rho/\rho$ and the peculiar velocity ${\bf v}$:
\begin{eqnarray} 
\fl
\dot \delta +3 {\cal H} (\omega_p -\omega)\delta - 3 \dot {\Psi}
(1+\omega)(1+R\delta) + (1+\omega) \nabla \cdot \left[(1+R\delta){\bf v}\right]
={{\cal C}_0 - \bar{\cal C}_0(1+\delta) \over\bar{\rho}} 
\nonumber\\ \nonumber\\
\fl
\dot{\bf v} +  \left[{\cal H}
\left(1-3\omega \right) + 
{1\over 1+R\delta}\left(
{({\cal C}_0 - \bar{\cal C}_0)R + \bar{\cal C}_0  \over \bar{\rho}} +
{\dot\omega + \dot{\omega}_p \delta \over 1 +\omega} + 
R  \dot\delta \right) \right]{\bf v} + 
\nonumber\\ 
~~~~~~~~~~~~~~~~~~~+{\nabla \delta p\over \bar{\rho}(1+\omega) (1+R\delta)} 
+ \nabla \Phi 
= -{{\bf {\cal C}} \over \bar{\rho} (1+\omega) (1+R\delta)}
\label{velocity} 
\end{eqnarray}
Here $R=(1+\omega_p) / (1+\omega)$ and $\nabla$ 
denotes the spatial gradient.

Notice that in the small perturbation limit ($\delta\rho/\rho,~
\delta p/ p << 1$), the parameter $\omega_p$ is related
to the sound speed $c_s^2$ in the fluid. In general, 
$\delta p = \delta p_a + \delta p_{na}$ where the two terms on the r.h.s are
the adiabatic and non--adiabatic contributions to the pressure perturbations
and $c_s^2$ is then their propagation speed in 
the fluid rest frame ($rf$), i.e. the frame where $T^0_i = 0$:
\begin{equation}
c_s^2= \left.{\delta p \over \delta\rho}\right|_{rf}
\label{cs}
\end{equation}
After performing a gauge transformation from the rest frame gauge 
to the conformal Newtonian gauge one arrives to the following 
relation between $\delta p$ and 
$\delta\rho$ \cite{kodama,hu,jussi} (we assume $c^2_s$ to depend
only by $\tau$):
\begin{equation}
\nabla \delta p = \nabla \delta p_a + \nabla \delta p_{na} =
c^2_{s_a} \nabla \delta \rho + (c^2_s-c^2_{s_a})[\nabla \delta \rho +
\dot {\bar{\rho}} {\bf v}]
\end{equation}
which can be rewritten as:
\begin{equation}
\nabla \delta p = 
c^2_{s} \nabla \delta \rho - 
(c^2_s-c^2_{s_a})[3 {\cal H}(1+\omega)\bar{\rho}- \bar{\cal C}_0] {\bf v}
\end{equation}
showing that the coupling enters $\delta p$ explicitly.
In the above relation, 
\begin{equation}
c^2_{s_a} = {\dot {\bar p} \over \dot
{\bar \rho}} = \omega + \dot{\omega} {\bar{\rho} \over \dot{\bar\rho}}
\end{equation}
is the adiabatic sound speed
which coincides with $c_s^2$ if the fluid is barotropic ($p=\omega(\rho)\rho$). 
Pure adiabatic perturbations, however, go unstable for 
fluids with $\omega<0$, e.g. DE fluids. 
Therefore, in order to ensure their 
stability ($c^2_{s}>0$) and have a DE component
phenomenologically viable, non--adiabatic modes must be also present 
(notice that the condition $c^2_s>0$ needs to be imposed by hand if DE  
is assumed to be a fluid while  $c_s^2=1$ follows without 
assumptions in the case of DE scalar fields. Also notice that a scalar 
field has in general a non--barotropic equation of state 
(see \cite{jussi,garriga} and next section)).
Nonetheless, non--adiabatic instabilities could arise in the presence of 
strong DM--DE couplings or on super--Hubble scales.
Adiabatic and non--adiabatic instabilities have been discussed in \cite{adiabatic}.

Equations (\ref{velocity}) can be generalized to the case in which anisotropic
stresses are present, i.e imperfect fluids. 
In that case we have $T^i_j = p \delta^i_j + \Pi^i_j$ where 
$\Pi^i_j$ is the anisotropic stress tensor which satisfies
$\Pi^\mu_\mu=0$ and $\Pi_\nu^\mu u^\nu = 0$. The equation for the peculiar 
velocity evolution 
must then be modified adding the term $\nabla\Pi$ to the l.h.s. 
(see \cite{ma} for further 
details).
In the next section we show that a scalar field minimally coupled to 
gravity has vanishing anisotropic stresses
and its stress energy tensor can be put in the perfect fluid form.
It follows that a DE scalar field can be described as a perfect fluid.
On the other hand, if one assumes DE to be a fluid, anisotropic stresses
can not be {\it a priori} excluded. Anyway, as we shall see shortly, the absence of  
anisotropic stresses leads $\Phi = \Psi$.

In additions to (\ref{velocity}), one has to consider
the equations of metric perturbations obtained from the perturbed part of 
the Einstein's equations. Because of the weak field approximation, the 
Einstein's tensor $G^\mu_\nu$ will be only linearly perturbed while the 
perturbations in the stress energy tensor components of each fluid $\alpha$ follow by 
subtracting their background values ${\bar T}^\mu_\nu$ from those obtained 
from (\ref{tensor}):
\begin{equation}
\delta G^\mu_\nu = 8\pi G \sum_\alpha\left(T^\mu_{(\alpha)\nu} - {\bar T}^\mu_{(\alpha)\nu} \right) 
\end{equation}
Einstein's equations then read:
\begin{eqnarray}
    \nabla^2\Psi - 3{\cal H} \left( \dot{\Psi} + {\cal H}\Phi
	\right) = 4\pi G a^2 \sum_\alpha\delta\rho_\alpha
	\nonumber\\ \nonumber\\
    \nabla^2 \left( \dot{\Psi} + {\cal H}\Phi \right)
	 = -4\pi G a^2 \partial_i \left[\sum_\alpha(\rho_\alpha +p_\alpha )v^i_\alpha\right] 
	 \nonumber\\ \nonumber\\
	     \ddot{\Psi} + {\cal H} (\dot{\Phi}+2\dot{\Psi})
	+\left(2{\ddot{a} \over a} - {\cal H}^2 \right)\Phi
	- { 1\over 3}\nabla^2 (\Psi-\Phi)
	= {4\pi\over 3} G a^2 \sum_\alpha\delta p_\alpha
        \nonumber\\ \nonumber\\
            \partial_i\partial_j(\Psi-\Phi) = 8\pi G a^2\sum_\alpha  
\Pi^i_{(\alpha)j}
\label{metricpert}
\end{eqnarray}
Thus, if anisotropic stresses are absent the last equation shows 
that $\Psi=\Phi$. In this case, focusing on scales well below the Hubble 
radius, we can neglect the second term on the l.h.s. in the first equation, 
because ${\cal H}\dot\Phi \sim {\cal H}^2 \Phi <<\nabla^2 \Phi$. This correspond
to the Newtonian limit and 
the first of equations (\ref{metricpert}) reduces to the usual 
Poisson equation: 
\begin{equation}
\nabla^2 \Phi
= 4\pi G \sum_\alpha \delta\rho_\alpha 
\end{equation}

In our particular case, neglecting the term proportional to $\dot \Psi$
in the first of (\ref{velocity}) because of the Newtonian approximation
and replacing the term 
$\dot \delta$ in the second, the perturbation equations
for our three perfect fluid components are:
\begin{eqnarray}
\label{b}
\nonumber\\
\fl
{\bf Baryons}~(\omega \approx 0, ~c^2_s ~~negligible  ~after ~recombination,
~ {\cal C}_\mu=0)
\nonumber\\
\nonumber\\ 
\fl
\dot \delta_{b} + \nabla \cdot \left[(1+\delta_{b}){\bf v}_{b} \right]
= 0\nonumber \\
\nonumber\\
\fl
\dot{\bf {v}}_b +  {\cal H}{\bf v}_{b} + \nabla \Phi = 0\\
\nonumber
\end{eqnarray}
\begin{eqnarray}
\label{dm}
\fl
{\bf DM ~(Pressureless ~perfect ~fluid)}~
(\omega=0,~\omega_p=0, ~{\cal C}_\mu=-C_\mu)\nonumber\\
\nonumber\\
\fl
\dot \delta_{dm} + \nabla \cdot \left[(1+\delta_{dm}){\bf v}_{dm} \right] = 
-{C_0 - {\bar C}_0 (1+\delta_{dm}) \over {\bar\rho}_{dm}}
\nonumber \\ 
\nonumber\\
\fl
\dot{\bf v}_{dm} +  \left( {\cal H}- { C_0\over {\bar\rho}_{dm}
(1+\delta_{dm})}\right)
{\bf v}_{dm} 
 + \nabla\Phi = {{\bf C} \over \bar{\rho}_{dm}  (1+\delta_{dm})}\\
\nonumber
\end{eqnarray}
\begin{eqnarray}
\label{de}
\fl
{\bf DE~(Perfect ~fluid ~with ~pressure)}~
(\omega={\bar p}_{de}/{\bar\rho}_{de},~\omega_p=\delta p_{de}/ 
\delta\rho_{de},~{\cal C}_\mu=C_\mu)\nonumber\\  
\nonumber\\
\fl
{\dot \delta}_{de} +3 {\cal H} (\omega_p -\omega)\delta_{de} 
+ (1+\omega) \nabla \cdot \left[(1+R\delta_{de}) {\bf v}_{de} \right]
= {C_0 - {\bar C}_0 (1+\delta_{dm}) \over {\bar\rho}_{de}}
\nonumber\\ 
\nonumber\\
\fl
\dot{\bf v}_{de} +  
\left[{\cal H}\left({1-3\omega+
(1-3\omega_p)R\delta_{de}\over 1+R\delta_{de}}\right)
 + {\dot\omega + \dot{\omega}_p \delta_{de} \over (1 +\omega) (1+R\delta_{de})}+
{( C_0 - \bar {C}_0)R + 
\bar{C}_0  \over \bar{\rho}_{de}(1+R\delta_{de})}
\right]{\bf v}_{de}+
 \nonumber\\
+{ \nabla \delta p_{de} \over \bar{\rho}_{de}(1+\omega)( 1+R\delta_{de})} 
+\nabla \Phi=-
{ {\bf C} \over {\bar\rho}_{de} (1+\omega) (1+R\delta_{de})}\\
\nonumber
\end{eqnarray}
Equations (\ref{b}),(\ref{dm}) and (\ref{de}) are valid to nonlinear order in the 
density contrast since we have not assumed $\delta<< 1$. They can be used to follow
the evolution of a collapsing region in nonlinear regime and could be
useful to study the clustering properties of DE by using the spherical collapse approach. 
The problem of spherical fluctuation growth in coupled DE models has been considered 
in \cite{mainini2}. 
There, equations (\ref{b}) and (\ref{dm}) were 
used to follow the evolution of a set of concentric shells of a spherical 
{\it top--hat} overdensity in baryons and DM while the DE was kept homogeneous.
 The shell description of the overdensity was needed because 
the coupling causes DM particles to have a different dynamics than the 
baryons. This can be seen from an inspection of the above equations. 
The most significant
result of that analysis is the high level of segregation between DM
and baryons arising during the growth of fluctuations: up to $\sim 60
\%$ of baryons can be expelled from the fluctuation in halo encounters
before its final virialization.

The same approach could be used relaxing the assumption of homogeneity for 
DE and including in the system an additional set of concentric shells for 
DE. Their evolution is then described by (\ref{de}).  
In contrast to DM, dynamics of DE shells is affected, not only by 
the coupling but also by the non--vanishing pressure of DE itself. We shall 
consider this problem in a subsequent paper.

A similar set of equations was used in \cite{abramo} to  study nonlinearities
in the case of uncoupled DE . 
However, the authors consider the pseudo--Newtonian approach of \cite{lima} 
and the DE equations they use lack of some terms with respect to (\ref{de}).

In the next section we show that  that equations (\ref{de}) applies 
equally well when DE is described by a scalar field.

\section{Scalar field}
\label{scalarfield}

In most DE models, DE is due to a scalar field either minimally or 
non--minimally coupled to matter and/or gravity.
In this paper we do not
discuss coupling to gravity, while coupling with matter will be considered 
only in a subsequent section. 
Anyway, results of this section are not affected 
by the presence of this latter kind of coupling and are fully general.

Here we show that the equations deduced in the previous section 
are valid also for a scalar field. 
In particular, following \cite{madsen}, we show that the stress energy tensor 
of a scalar field, if minimally coupled to gravity, can be put 
in the perfect fluid form.
As a consequence equations (\ref{velocity}) as well as 
(\ref{de}) can be applied.  

Let us start with a minimally coupled scalar field $\phi$ self--interacting 
through a potential $V(\phi)$.
In a spacetime described by a metric $g_{\mu\nu}$ with signature $(-+++)$
its Lagrangian reads:
\begin{equation}
{\cal L} =  \sqrt{-g} \left[-{1\over 2} \partial_\mu \phi \partial^\mu \phi
- V(\phi) \right]
\label{lagrangian}
\end{equation}
and the stress energy tensor takes the form:
\begin{equation}
T^\mu_\nu = \partial^\mu \phi \partial_\nu \phi - \delta^\mu_\nu \left(
{1\over 2} \partial_\sigma \phi \partial^\sigma \phi +V\right)  
\label{phitensor}
\end{equation}
The equations of motion then follow either from $\nabla_\mu T^\mu_\nu = 0$ 
or from the Euler--Lagrange equation:
\begin{equation}
\left[
{1\over \sqrt{-g}} \partial_\mu 
\left( \sqrt{-g} \partial^\mu \right)\right]     \phi -  V'  =0
\label{motion}
\end{equation}
(the prime denotes the derivative with respect to $\phi$).

Assuming that the vector $\partial^\mu \phi$ is timelike, namely 
$\partial^\mu \phi \partial_\mu \phi < 0$, one can assign a 4--velocity $u^\mu$
to the scalar field.
It can be chosen as the unique timelike 
vector with unit magnitude constructed from $\partial^\mu \phi$
(for more details see \cite{madsen,bruni}):
\begin{equation}
u^\mu = -
{\partial^\mu \phi \over \sqrt{-\partial_\sigma \phi \partial^\sigma \phi}} 
\\
u^\mu u_\mu = -1
\label{field4vel}
\end{equation}

From this choice of the 4--velocity, it follows 
that the scalar field $\phi$ can be represented as a perfect
fluid and the stress energy tensor (\ref{phitensor}) can be put in the form 
(\ref{stressenergy}).
In order to show this, let us start from the stress energy 
tensor of a general fluid: 
\begin{equation}
T_{\mu \nu} = p g_{\mu \nu} + (\rho + p) u_{\mu} u_{\nu} + 
q_\mu u_\nu + u_\mu q_\nu +\Sigma_{\mu\nu}
\label{general}
\end{equation}
where possible anisotropic stresses are accounted for by the heat flux vector 
$q^\mu$  and the tensor $\Sigma_{\mu\nu}$ which satisfy
$u_\mu q^\mu=0$, $\Sigma^\mu_\mu=0$ and $\Sigma_\nu^\mu u^\nu = 0$.
We show now that for a scalar field these terms vanish.
Making use of the projection tensor $h_{\mu\nu} = g_{\mu\nu} +  u_{\mu} u_{\nu}$,
one can extract from (\ref{general}) all the relevant quantities:
\begin{equation}
\fl
\rho=T_{\mu \nu} u^{\mu} u^{\nu}\\
p={1\over3}T_{\mu \nu}h^{\mu \sigma}h^\nu_\sigma\\
q_\mu =  -T_{\sigma \lambda}u^{\sigma}h^\lambda_\mu\\
\Sigma_{\mu\nu}= T_{\sigma \lambda}h^\sigma_\mu h^\lambda_\nu -p h_{\mu\nu}
\end{equation}
By using (\ref{phitensor}) and  (\ref{field4vel}), the above quantities can be found for the
scalar field $\phi$:
\begin{equation}
\fl
\rho=-{1\over2}\partial_\mu \phi \partial^\mu \phi + V\\
p=-{1\over2}\partial_\mu \phi \partial^\mu \phi - V\\
q_\mu=0\\
\Sigma_{\mu\nu}= 0
\end{equation}
Both the heat flux and the anisotropic stress tensor vanish, showing that 
the scalar field stress energy tensor has the usual perfect fluid form.
 
However, it is important to mention that, although this mathematical equivalence relates a scalar field with perfect fluid models, 
differences remain in their physical interpretation.  In general, a fluid is a continuum model deduced from thermodynamical considerations of a system containing a significant number of particles interacting through elastic collisions. Its continuum variables (e.g., $\rho$, $u^\mu$) are defined in terms of fundamental discrete variables in a limit procedure. Once the large particle number limit
is violated the model is expected to fail. Because of the thermodynamical motivation, physical constraints on $\rho$ and $u^\mu$ are then imposed, namely that $\rho>0$ and that $u^\mu$ is timelike.
On the other hand, a scalar field model is a continuum model and the solutions of its equations
are generally not subjected to the constraints above mentioned. Thus, the assumption $\partial^\mu\phi\partial_\mu\phi<0$  only refers to those scalar field solutions that can be interpreted as physical fluids.

Let us now consider a perturbed FRW Universe described by the metric (\ref{metric}).
Equation of motion (\ref{lagrangian}) then reads:
\begin{equation}
\fl
\ddot\phi + \left(2{\cal H} -\dot \Phi -3\dot \Psi\right)\dot\phi -
(1+2\Psi+2\Phi) \nabla^2 \phi+a^2 (1+2\Phi)V^\prime =
\nabla(\Phi-\Psi) \cdot \nabla \phi
\label{phieq}
\end{equation}
and the components of the fully nonlinear stress energy tensor are:
\begin{eqnarray}
T^0_0 = -{{\dot\phi}^2 \over 2a^2}(1-2\Phi) + 
{\sum_i(\partial_i \phi)^2 \over 2a^2}(1+2\Psi) - V 
\nonumber\\ \nonumber\\
T^0_i = -{\dot \phi \over a^2}
(1-2\Phi) 
\partial_i \phi
\nonumber\\ \nonumber\\
T^i_j ={\partial_i \phi \partial_j \phi \over a^2}(1+2\Psi)+ \delta^i_j \left[{\dot {\phi}^2 \over 2a^2}(1-2\Phi) - 
{\sum_n(\partial_n \phi)^2 \over 2a^2}(1+2\Psi) - V\right]
\label{tensor2}
\end{eqnarray} 
Then, decomposing $\phi$ as the sum of an unperturbed part $\bar \phi$
and a perturbed one $\delta \phi$:
\begin{equation}
\phi={\bar\phi}(\tau) + \delta\phi(\tau,x^i)
\label{campo}
\end{equation}
the 0th--order terms of (\ref{phieq}) give the equation for 
the homogeneous part of $\phi$:
\begin{equation}
\ddot {\bar\phi} + 2{\dot a\over a}\dot{\bar\phi} 
+a^2 {\bar V}^\prime =0
\label{}
\end{equation}
while, assuming no anisotropic stresses ($\Phi=\Psi$), 
the higher order terms yield for the perturbations:
\begin{eqnarray}
\fl
\ddot{\delta\phi} + 
\left(2{\cal H} -4\dot \Phi\right)\dot{\delta\phi} -
4\dot \Phi\dot{\bar\phi}-
(1+4\Phi) \nabla^2 \delta\phi+ 
a^2\left[(1+2\Phi) V^\prime 
-{\bar V}^\prime \right] =0
\label{perturbations}
\end{eqnarray}
From (\ref{tensor2}) one can also work out the background energy density 
and pressure and the corresponding perturbations:
\begin{eqnarray}
\fl
{\bar\rho} = {{\dot {\bar\phi^2}} \over 2a^2}+{\bar V}~~~~~~~
\delta\rho = -{\dot  {\bar\phi^2} \over a^2}\Phi + 
{\dot{\bar\phi}\dot{\delta\phi}\over a^2}(1-2\Phi) +
{\dot{\delta\phi}^2\over 2a^2}
(1-2\Phi)-{\sum_i(\partial_i\delta\phi)^2 \over 2a^2}(1+2\Psi) + \delta V
\nonumber\\
\nonumber\\
\fl
{\bar p} = {{\dot  {\bar\phi^2}} \over 2a^2}-{\bar V}~~~~~~~
\delta p = -{\dot  {\bar\phi^2}\over a^2}\Phi + 
{\dot {\bar\phi}\dot{\delta\phi}\over a^2}(1-2\Phi) 
+{\dot{\delta\phi}^2\over 2a^2}
(1-2\Phi)-{\sum_i(\partial_i \delta\phi)^2 \over 2a^2}(1+2\Psi) - \delta V
\nonumber\\
\end{eqnarray}
where $\delta V = V - {\bar V}$ and ${\bar V} = V({\bar\phi})$. The scalar 
field peculiar velocity $v^i$ is obtained by comparing the expression 
(\ref{field4vel}), which can be rewritten as:
\begin{equation}
u^\mu=  a^{-1}\left[(1+2\Phi)-(1-2\Psi)\sum_j 
\left({\partial^j \phi
\over \partial^0 \phi}\right)^2
 \right]^{-{1\over 2}}{\partial^\mu \phi
\over \partial^0 \phi} 
\end{equation}
 with the fluid 4--velocity:
\begin{equation}
u^\mu={dx^\mu \over \sqrt{-ds^2}}=  a^{-1}\left[(1+2\Phi)-(1-2\Psi)
\sum_j \left({dx^j \over d\tau}\right)^2 \right]^{-{1\over 2}}
{dx^\mu \over d\tau}
\end{equation}
Taking $\mu=i$, it then follows:
\begin{equation}
v^i={dx^i \over d\tau}= {\partial^i \phi
\over \partial^0 \phi}
\end{equation}
As we are interested in the small scale and weak gravitational field limits
some approximations are possible.
Terms proportional to $\Phi$ can be dropped in equation (\ref{perturbations}).
We can also neglect the time derivatives of the 
gravitational potential with respect to $\cal H$, 
since $\dot\Phi\sim \Phi {\cal H}$. We are then left with:
\begin{equation}
\ddot{\delta\phi} + 
2{\cal H}\dot{\delta\phi} -
\nabla^2 \delta\phi +a^2\delta V^\prime=
  4\dot \Phi\dot{\bar\phi}
\label{deltaphieq}
\end{equation}

\begin{table}
\caption{\label{table} Scalar field--perfect fluid correspondence 
in the small velocity and weak gravitational field limits}
\begin{indented}
\item[]\begin{tabular}{@{}llll}
\br
   &{\bf Perfect fluid}&{\bf Scalar field}\\
\mr
\\
$u^0 ~~$&$a^{-1}(1-\Phi)$&$a^{-1}(1-\Phi)$\\ \\
$u^i ~~$&${dx^i \over d\tau}a^{-1}$&${\partial^i \phi \over \partial^0 \phi}a^{-1}$\\ \\
$v^i ~~$ &${dx^i \over d\tau}$&${\partial^i \phi \over \partial^0 \phi}$\\ \\
$T^0_0 ~~$&$-\rho$&$-{\dot\phi^2\over 2a^2}(1-2\Phi)-V(\phi)$\\ \\
$T^i_0 ~~$&$-(\rho+p)v^i$&$\dot\phi\partial^i \phi$\\ \\
$T^i_j ~~$&$\delta^i_j p$&
$\delta^i_j\left[{\dot\phi^2\over 2a^2}(1-2\Phi)-V(\phi)\right]$\\ \\
\br
\end{tabular}
\end{indented}
\end{table}

The correspondence between the perfect fluid and scalar field descriptions 
is summarized in the table \ref{table} where
the expressions taken by the 4--velocity and the stress
energy tensor components in the small velocity and weak gravitational 
field limits are given.
Notice that these limits, in the 
scalar field description, correspond to considering only linear terms in 
${\partial^\mu \phi / \partial^0 \phi}$, $\Psi$ and $\Phi$ and neglecting 
terms involving products of the former and the metric perturbations.
It is worth to mentioning that equation (\ref{deltaphieq}) 
is equivalent to the first of (\ref{de}), from which it can be directly derived
by setting $C_\mu=0$.
However, once the prescriptions of Table \ref{table} are taken into account,
equation (\ref{deltaphieq}) is more simply deduced from (\ref{conservation})
or directly from (\ref{eq1}) after considering the small scale limit and 
eliminating the homogeneous part. 

As mentioned in the previous section, in general, a scalar field  
cannot be considered as a barotropic fluid being its equation of state 
given by:
\begin{equation}
p=\rho-2V
\end{equation}
Therefore, for small perturbations, if $ V \ne const$, we have 
$\delta p / \delta \rho \ne \dot{\bar p}/\dot{\bar\rho} = c^2_{s_a}$
yielding also non--adiabatic contributions to the fluctuations. 
The sound speed is given again by (\ref{cs}) where the rest frame is defined 
by the hypersurfaces $\phi=const$ orthogonal to the rest frame 4-velocity 
(\ref{field4vel}).
By definition, in the rest frame, the scalar field carries no perturbations 
($\left.{\delta\phi} \right|_{rf} = 0$) leading 
$\left.{\delta V} \right|_{rf} = 0$. 
It follows that
energy density and pressure perturbations come purely from 
kinetic terms so that 
$\left.{\delta \rho} \right|_{rf}= \left.{\delta p}\right|_{rf}$ and
$c^2_s=1$ independent of the form of $V(\phi)$ (for a detailed discussion 
see \cite{langlois}).

In the next section we introduce the coupling through a discussion on the
coupled DE model.

\section{Coupled Dark Energy}
\label{example}
In addition to self--interaction, a scalar field can in principle be 
coupled to any other field present in nature. However, if this field 
is the one which accounts for DE, in order to drive the cosmic acceleration, 
its present time mass is expected to be, at least on large scales,  
$m_\phi \sim H_0 \sim 10^{-33} eV$.
Such a tiny mass gives rise to long--range interactions which could be tested
with fifth--force type experiments. 
Couplings to ordinary particles are strongly constrained by such a kind of 
experiments but limits on the DM coupling are looser (constraints on coupling
 for specific models were obtained in \cite{maccio,constraints} from CMB,
 N-body 
simulations and matter power spectrum analysis).  
If present, DM coupling could have a relevant role in 
the cosmological evolution affecting not only the overall cosmic expansion
but also modifying the DM particles dynamics with  relevant 
consequences on the growth of the density perturbations in both 
linear and nonlinear regime (e.g., on halo density profiles, 
mass function and its evolution)  \cite{maccio,amendola,dolag}.
Here we review one of the most popular models where a coupling between DM and 
DE is present, namely coupled DE \cite{amendola}, and show how the formalism 
introduced in Sec. \ref{perturbation} can be applied.

The model is quite general and includes a vast class of models also motivated 
by string theory as proposed in \cite{interaction,interaction2}. 
Conservation equations for 
baryons, DM and a DE scalar field $\phi$ read:
\begin{equation}
\fl
\nabla_{\mu} T^{\mu}_{(b) \nu} = 0\\
\nabla_{\mu} T^{\mu}_{(dm) \nu} = -c(\phi)T_{(dm)}\nabla_\nu\phi\\
\nabla_{\mu} T^{\mu}_{(\phi) \nu} = c(\phi)T_{(dm)}\nabla_\nu\phi
\label{continuity2}
\end{equation}
so that the coupling  $C_\mu$ introduced in Sec. \ref{background} 
now takes the form:
\begin{equation}
C_\mu = c(\phi)T_{(dm)}\nabla_\mu\phi
\label{coupling}
\end{equation}
where $T_{(dm)}$ is the trace of the DM stress energy tensor and $c(\phi)$
parametrizes the strength of the DM--DE interaction. Such a coupling can be
derived from a scalar--tensor theory after performing a conformal 
transformation (see \cite{fujii} for a review on scalar--tensor gravity). 
Different couplings were also considered by various 
authors \cite{zimdahl}. 

The equation of motion of the scalar field follows from  
the last of (\ref{continuity2}) and (\ref{phitensor}) or 
equivalently from the lagrangian
\begin{equation}
{\cal L} = \sqrt{-g} \left[-{1\over 2} \partial_\mu \phi \partial^\mu \phi
- V(\phi)  - B(\phi)m{\bar\chi}\chi\right]
\label{lagrangian2}
\end{equation}
in the case when $c(\phi)= d(ln B)/d\phi$.
The last term in (\ref{lagrangian2}) sets the coupling between the 
DE and the spinor field
$\chi$ supposed here to yield DM (results are however independent of whether the 
DM particles are scalars or fermions). Notice that 
in such a theory the mass of the DM particles, $m_{dm}=B(\phi)m$,
as well as their energy density $\rho_{dm} = m_{dm} <{\bar\chi}\chi>$
are $\phi$--dependent. 
In this case equation (\ref{deltaphieq}) reads:
\begin{equation}
\ddot{\delta\phi} + 
2{\cal H}\dot{\delta\phi} -
\nabla^2 \delta\phi +a^2\delta V_{eff}^\prime=
  4\dot \Phi\dot{\bar\phi}
\label{deltaphieq2}
\end{equation}
where $\delta V_{eff} = V_{eff} - {\bar V}_{eff}$ and $V_{eff} = V + \rho_{dm}$.
Notice that no assumptions obout the value of $\delta\phi$ were made in deriving 
the above equation which is valid also for nonlinear $\delta\phi$.

It is also worth to mentioning that, unlike the uncoupled case, 
the effective mass of the scalar field,
$m_{\phi}=V_{eff}^{\prime\prime}(\phi)=
V^{\prime\prime}(\phi)+c^{\prime}(\phi)\rho_{dm}+c^2(\phi)\rho_{dm}$,
as well as the value of $\phi$
depend on the local DM density $\rho_{dm}$. It follows that, 
in massive objects, 
where the DM density is high compared 
to the background density, 
$m_\phi$ could be very different from the value it takes in the cosmos
where $\rho_{dm} \approx \rho_{cr}$ ($\rho_{cr}$ being the critical energy 
density) and $\phi=\bar{\phi}$. Thus, perturbations in the DE energy 
density and pressure could be non--negligible on the scales relevant for 
the structure formation.
Effects of a scale--dependent mass was considered in the contest of the 
so--called {\it chameleon theory} \cite{chameleon} where, unlike the coupled DE
case, the scalar field universally couples to all the kinds 
of matter, e.g. DM and baryons. However, even though this theory simultaneously 
provides a viable cosmology and  a
mechanism which allows the scalar field to evade constraints 
from fifth force effects, a very low value of the energy 
scale entering the chameleon potential is needed.
Although tracker potentials were considered,
it seems therefore hard to avoid fine--tuning and coincidence problems.

Alternatively to the equation (\ref{deltaphieq2}), in order to study 
the evolution of DE perturbations, one can use the perfect fluid description
and the equations (\ref{de}) once the coupling vector $C_\mu$ is rewritten in
term of $\rho_{de}$, $p_{de}$ and ${\bf v}_{de}$. 
Let us consider first the case when the parameter $c$ of (\ref{coupling}) is 
constant. In this case, by using the relations of table \ref{table}, 
the coupling vector becomes: 
\begin{eqnarray}
C_0=c\bar{\rho}_{dm}(1+\delta_{dm})
[\bar{\rho}_{de}(1+\omega)(1+R\delta_{de})]^{1\over2}a 
\nonumber\\ \nonumber\\
{\bf C}=-c\bar{\rho}_{dm}(1+\delta_{dm})
[\bar{\rho}_{de}(1+\omega)(1+R\delta_{de})]^{1\over2}a {\bf v}_{de}
\end{eqnarray}

Problems can arise when $c$ is $\phi$ dependent. However, if the potential $V(\phi)$ is a strictly monotonic function, as in the case of an inverse power law or an exponential potential, 
one can invert it to obtain $\phi$ as function of $V=(\rho_{de}-p_{de})/2$. As an example, consider the Ratra-Peebles potential $
V=\Lambda^{4+\alpha}/ \phi^\alpha$ where  $\alpha>0$. In this case,
\begin{equation}
\phi=\left({\Lambda^{4+\alpha} \over V}\right)^{1\over \alpha}=
\left({2\Lambda^{4+\alpha} \over
 \bar{\rho}_{de}[(1-\omega)+\delta_{de}(1-\omega_p)]}\right)^{1\over \alpha}
\end{equation}
so that $c=c(\rho_{de}-p_{de};\Lambda,\alpha)$.  Notice that, once the energy scale $\Lambda$ is fixed,
the $\alpha$ parameter follows by requiring $\bar{\rho}_{de,0}$ (or $\bar{\rho}_{m,0}=\bar{\rho}_{dm,0}
+\bar{\rho}_{b,0}$) to have a value as inferred by observations (here the subscript $_0$ refers to the present time). Alternatively, one can fix $\alpha$
and $\bar{\rho}_{de,0}$ (or $\bar{\rho}_{m,0}$) then find $\Lambda$.

\section{Conclusions}
\label{conclusion}

One of the main topics which has attracted a deep interest in recent years 
concerns the clustering properties of dynamical DE, in particular 
when a coupling with matter is present.
Scalar fields are among the favorite candidates for DE.
However, in certain circumstances, it can be more advantageous describing 
DE in terms of a perfect fluid rather than in terms of scalar fields.

This paper is a technical review meant to set up a formalism for the study of 
nonlinearity in the presence of DM--DE coupling. To this aim, 
we derive the Newtonian limit of the perturbation equations for DM and DE 
(and baryons) valid to nonlinear order in the density contrast,
highlighting the equivalence between the perfect fluid and the scalar field 
descriptions of DE. We then provide some prescriptions 
(summarized in table \ref{table}) to relate the two 
descriptions and apply them to the 
coupled DE case. This enables us to give the expression
of the coupling vector in terms of the fluid variables and
show that, in the presence of coupling, the mass of DE quanta could increase
where large DM condensations are present. As a consequence, 
also DE can take part in the clustering process.

The framework here presented gives the DE equations of motion in 
the presence of coupling once the time evolution of the DE equation of state, 
$\omega(a)$, is assigned, without reconstructing the self--interacting 
potential $V(\phi)$. It can be useful to follow the evolution of a 
collapsing region in nonlinear regime in order 
to study the clustering properties of DE, e.g. by using the spherical 
collapse approach. This point is currently under investigation.

\begin{ack}
I wish to thank A. Gardini, D. Mota, D. Puetzfeld, {\O}. Elgar{\o}y and 
L. Amendola for useful hints and discussions.  
I also thank S. Bonometto for carefully reading the manuscript, helpful
comments and suggestions. 
This work is supported by the Research Council of Norway, project number 162830.
\end{ack}

{}


\section*{References}
\begin{thebibliography}{10}

\bibitem{tegmark}
 Tegmark M. et al., 2004, Phys.Rev. D69, 10350;   De Bernardis et al.,
 2000 Nature 404, 955;
 Hanany S. et al, 2000, ApJ 545, L5; Halverson N.W. et al. 2002, ApJ 568, 38;
Percival W.J. et al., 2002, MNRAS, 337, 1068
 , Riess, A.G. et al., 1998, Aj 116, 1009;
 Perlmutter S. et al., 1999, Apj, 517, 565

\bibitem{oystein}
Kristiansen J. R., Elgar{\o}y {\O}., Dahle H., 2007, Phys. Rev. D75, 083510;
see also Elgar{\o}y {\O}. et al., 2002, Phys. Rev. Lett. 89, 061301

\bibitem{copeland}
Copeland E. J., Sami M., Tsujikawa S., 2006, Int.J.Mod.Phys D15, 1753

\bibitem{peebles}
Peebles P.J.E. \& Ratra B., 2003, Rev.Mod.Phys. 75, 559

\bibitem{mainini}
Wang L. \& Steinhardt P.J., 1998, ApJ, 508, 483;
Mainini R., Maccio' A., Bonometto S., 2003, New Astron. 8, 173;
Mainini R., Maccio' A., Bonometto S., Klypin A., 2003, ApJ. 599, 24;
Klypin A., Maccio' A., Mainini R., S.A. Bonometto, 2003, ApJ, 5999, 24;
Lokas E. L., Bode P., Hoffman Y., 2004, MNRAS, 349, 595;
Horellou C., Berge J., 2005, MNRAS, 360, 1393


\bibitem{interaction}
Ellis J., Kalara S., Olive K.A. \& Wetterich C., 1989, Phys. Lett. B228, 264; 
Wetterich C., 1995, A\&A 301, 321 
Amendola L., 1999, Phys.Rev. D60, 043501
Gasperini M., Piazza F.\& Veneziano G., 2002, Phys.Rev. D65, 023508

\bibitem{interaction2}
Casas J.A., García--Bellido J \& Quiros M., 1992, Class.Quant.Grav. 9, 1371;
Anderson G.W . \& Carroll S.M., Procs. of ``COSMO-97, First International 
Workshop on Particle Physics 
and the Early Universe'', Ambleside, England, September 15-19, 1997, 
astro-ph/9711288;
Bartolo N. \& Pietroni M., 2000, Phys.Rev. D61, 023518; 
Mangano G., Miele G., Pettorino V., 2003, Mod.Phys.Lett. A18, 831
Pietroni M., 2003, Phys.Rev D67, 103523; 
Chimento L.P., Jakubi A.S., Pavon D. \& Zimdahl W.,2003, Phys.Rev D67, 083513;
Rhodes C.S., van de Bruck C, Brax P., \& Davis A.C., 2003, Phys.Rev. D68, 
083511;
Farrar G.R. \& Peebles P.J.E., 2004, ApJ 604, 1
Gromov A., Baryshev Y. \& Teerikorpi P., 2004, A\&A, 415, 813

\bibitem{matarrese}
Matarrese S., Pietroni M., Schimd C., 2003, JCAP 0308, 005;
Perrotta F., Matarrese S., Pietroni, Schimd C., 2004, Phys.Rev. D69, 084004

\bibitem{nunes}
Nunes N. J., da Silva A. C., Aghanim N., 2005,  A\&A 450, 899;
Mota D. \& van de Bruck C., 2004, A\&A, 421,71;
Maor I., Lahav O., 2005, JCAP 0507, 003;
Wang P., 2006, ApJ 640,18; 
Nunes J. N. \& Mota D., 2006,  MNRAS 368, 75; 
Manera M. \& Mota D., 2006, MNRAS 371, 1373;
Dutta S. \& Maor I., 2007, Phys.Rev. D75, 063507

\bibitem{mota}
Mota D., Shaw J., Silk J., 2008, ApJ 675, 29

\bibitem{madsen}
Madsen M. S., 1988, Class.Quant.Grav., 5, 627

\bibitem{bruni}
Bruni M., Ellis G. F. R., Dunsby K. S., 1992, 921
 
\bibitem{ma}
Ma C. P. \& Bertschinger E., 1995, ApJ 455, 7

\bibitem{kodama}
Kodama H. \& Sasaki M.,1984, Prog.Theor.Phys.Suppl. 78, 1

\bibitem{bert}
Bertschinger E., 1998, ARAA, 36, 599

\bibitem{spheric1}
Gunn J. \& Gott J.R , 1972, ApJ, 176, 1G;
Gott R. \& Rees M., 1975, A\&A,  45, 365G

\bibitem{peebles2}
Peebles P. J. E., 1980, {\it Large-Scale Structure of the Universe}
(Princeton Univ. Press)

\bibitem{spheric2}
Lahav O., Lilje P.R., Primack J.R. \& Rees M., 1991, MNRAS, 282, 263;
Eke V.R., Cole, S. \& Frenk, C.S., 1996, MNRAS, 282, 263;
Brian G. \& Norman M., 1998, ApJ, 495, 80

\bibitem{mainini2}
Mainini R., 2005, Phys.Rev. D72, 083514;
Mainini R. \& Bonometto S. A., 2006, Phys.Rev. D74, 043504


\bibitem{PS}
Press W.H. \& Schechter P.,  1974, ApJ, 187, 425

\bibitem{sheth}
Sheth R.K. \& Tormen G., 1999, MNRAS, 308, 119;
Sheth R.K., Mo H.J. \& Tormen G., 2001, MNRAS, 323 ,1;
Sheth R.K. \& Tormen G., 2002, MNRAS, 329, 61;
Jenkins, A., Frenk C.S., White S.D.M., Colberg J.M.,
Cole S., Evrard A.E., Couchman H.M.P. \& Yoshida N., 2001, MNRAS, 321, 372

\bibitem{lacey}
Lacey C. \& Cole S., 1993, MNRAS, 262, 627;
Lacey C. \& Cole S., 1993, MNRAS, 271, 676

\bibitem{hu}
HU W., 1998, ApJ, 506, 485

\bibitem{jussi}
V$\ddot{a}$liviita J., Majerotto E., Maartens R., 2008, arXiv:0804.0232 

\bibitem{garriga}
Garriga J, Mukhanov V. F., 1999, Phys.Lett.B 458, 219

\bibitem{adiabatic}
Koivisto T., 2005, Phys.Rev.72, 043516;
Sandvik H., Tegmark M., Zaldarriaga M., Waga I., 2004, Phys.Rev. D69, 123524;
Afshordi N., Zaldarriaga M., Kohri K., 2005, Phys.Rev. D72, 065024; 
Kaplinghat M. \& Rajaraman A., 2007 Phys.Rev. D75, 103504; 
Bjaelde O. E., Brookfield A. W., van de Bruck C., Hannestad S.,
Mota D., Schrempp L., Tocchini--Valentini D., 2008, JCAP, 0801, 026;
Bean R., Flanagan E. E., Trodden M., 2008, New J. Phys., 10, 033006; 
Avelino P. P., Beca L. M. G., Martins C. J. A., 2008, Phys. Rev. D 77, 063515;
Ichiki K. \& Keum Y. Y., 2008, arXiv:0803.3142;


\bibitem{abramo}
Abramo L. R., Batista R. C., Liberato L., Rosenfeld R.,2007, JCAP 11, 012

\bibitem{lima}
Lima J. A. S., Zanchin V., Brandenberger R. H., 1997, MNRAS 291, L1

\bibitem{langlois}
Langlois D. \& Vernizzi F., 2007, JCAP, 0702, 017;
Tsagas C. G., Challinor A., Maartens R., arXiv:0705.4397

\bibitem{maccio}
Maccio' A. V., Quercellini C., Mainini R., Amendola L., Bonometto S. A., 
2004, Phys. Rev. D69, 123516

\bibitem{constraints}
Amendola L. \& Quercellini C., 2003, Phys. Rev. D69;
Olivares G., Atrio--Barandela F., Pavon D., 2005, Phys.Rev. D71, 063523;
Lee S., Liu G. \&  Ng K., 2006, Phys.Rev. D73, 083516;
Guo Z., Ohta N. \& Tsujikawa S., 2007, Phys.Rev. D76, 023508; 
Mainini R. \& Bonometto S. A., 2007, JCAP 06,020

\bibitem{amendola}
Amendola L., 2000, Phys.Rev. D62, 043511;
Amendola L., 2004, Phys.Rev. D69, 103524

\bibitem{dolag}
Dolag K. et al., 2004,  A\&A 416, 853;
Olivares G., Atrio--Barandela F., Pavon D., 2006, Phys.Rev. D74, 043521

\bibitem{fujii}
Fujii J. \& Maeda K., 2003, {\it The Scalar-Tensor Theory of Gravitation},
Cambridge University Press;
Faraoni V., 2004, {\it Cosmology in Scalar-Tensor Gravity},
Kluwer Academic Pub
 
\bibitem{zimdahl}
Zimdahl W. \& Pavon D, 2003, Gen.Rel.Grav. 35, 413;
Cai R. \& Wang A., 2005, JCAP 0503, 002;
Chimento L. P., Jakubi A. S.,  Pavon D., Zimdahl W., 2003, Phys.Rev. D67, 
083513

\bibitem{chameleon}
Khoury J. \& Weltman A., 2004, Phys.Rev. D69, 044026;
Brax P., van de Bruck C., Davis A. C., Khoury J., Weltman A., 2004, 
Phys.Rev. D70, 123518




\end{thebibliography}
\end{document}